\documentclass[prl,reprint,floatfix,showpacs,superscriptaddress,twocolumn]{revtex4-1}
\usepackage[USenglish]{babel}
\usepackage{amsmath,amssymb,amsfonts}
\usepackage{epsfig}
\usepackage{subfigure}

\usepackage[colorlinks=true,linkcolor=blue,citecolor=red]{hyperref}

\newcommand{\eq}[2]
{
  \begin{equation}
    #1
    \label{#2}
  \end{equation}
}

\newcommand{\eqnn}[1]
{\begin{equation*}
    #1
  \end{equation*}}

\newcommand{\eqsplit}[2]
{
  \begin{equation}
    \begin{split}
      #1
    \end{split}
    \label{#2}
  \end{equation}
}

\newcommand{\figu}[1]
{Fig.\ref{#1}}

\newcount\colveccount
\newcommand*\colvec[1]{
  \global\colveccount#1
  \begin{pmatrix}
    \colvecnext
  }
  \def\colvecnext#1{
    #1
    \global\advance\colveccount-1
    \ifnum\colveccount>0
    \\
    \expandafter\colvecnext
    \else
  \end{pmatrix}
  \fi
}

\newtoks\rowvectoks
\newcommand{\rowvec}[2]{%
  \rowvectoks={#2}\count255=#1\relax
  \advance\count255 by -1
  \rowvecnexta}
\newcommand{\rowvecnexta}{%
  \ifnum\count255>0
  \expandafter\rowvecnextb
  \else
  \begin{pmatrix}\the\rowvectoks\end{pmatrix}
  \fi}
\newcommand\rowvecnextb[1]{%
  \rowvectoks=\expandafter{\the\rowvectoks&#1}%
  \advance\count255 by -1
  \rowvecnexta
}

\def\bcen{\begin{center}}
\def\ecen{\end{center}}

\def\a{\alpha}

\def\e{\varepsilon}

 \def\s{\sigma}

\def\o{\omega}

\def\HH{{\cal H}}

\def\GG{{\cal G}}

\def\=={\equiv}

\def\qed{\raise1pt\hbox{\vrule height5pt width5pt depth0pt}}
\def\iome{i\omega_n}

\def\cG0{{\cal G}_0} 
\def\cG{{\cal G}}

\def\up{\uparrow} 
 
\def\dw{\downarrow}
\def\bra{\langle} 
\def\ket{\rangle}
\def\ka{{\bf k}}

 \def\Im{\mbox{Im}}

\def\=={\equiv}

\def\Im{{\rm Im}} 
\def\Re{{\rm Re}}

\def\ep0{\epsilon_{p}} 
\def\ed0{\epsilon_{d}}

\def\v0{t_{pd}} 
\def\tpd{v_\mathrm{0}}

\def\nd{\langle n_d\rangle}
\def\np{\langle n_p\rangle}

\def\pamrowvec{\rowvec{2}{p_{\ka\s}}{d_{\ka\s}}}

\begin{document}
\title{Mott transitions with partially-filled correlated orbitals.}

\author{A.~Amaricci}
\affiliation{Scuola Internazionale Superiore di Studi Avanzati (SISSA)
  and Consiglio Nazionale delle
  Ricerche, Istituto Officina dei Materiali (IOM), Via Bonomea 265,
  34136 Trieste, Italy}

\author{L.~de'~Medici}
\affiliation{European Synchrotron Radiation Facility, 
BP 220, F-38043 
Grenoble Cedex 9, France}

\author{M.~Capone}
\affiliation{Scuola Internazionale Superiore di Studi Avanzati
  (SISSA), Via Bonomea 265, 34136 Trieste, Italy}


\begin{abstract}
We investigate the metal-insulator Mott transition in a generalized version
of the periodic Anderson model, in which a band of itinerant electrons is hybridrized 
with a narrow  and strongly correlated band. 
Using dynamical mean-field theory, we show that the precondition for a 
Mott transition is an integer total filling of the two bands, while for an integer
constant occupation of the correlated band the system remains 
a correlated metal at arbitrary large interaction strength. 
We picture the transition at a non-integer filling of the 
correlated orbital as the Mott localization of the singlet states
between itinerant and strongly interacting electrons, having
occupation of one per lattice site.  We show that the Mott transition is of
the first-order and we characterize the nature of the resulting
insulating state with respect to relevant physical parameters, such as the charge-transfer energy. 
\end{abstract}


\maketitle

\paragraph{Introduction. -- }
Mott localization induced by strong repulsive interaction is intrinsically a physics of commensuration. 
A paradigmatic example is the 
Mott-Hubbard transition in the Hubbard model which, in
the absence of symmetry breaking, only takes place if the number of 
correlated electrons equals the number of lattice
sites for a single-orbital model~\cite{Mott1968RMP,Imada1998RMP,Kotliar1999EPJB,Kotliar2000PRL},
or an integer multiple of it in the multi-orbital
case~\cite{Kotliar1996PRB,Rozenberg1997PRB,Koga2005PBCM,Kotliar2006RMP}.
A metal-insulator transition (MIT) occurs as the local repulsion reaches a critical value of the
order of the bandwidth, when it becomes energetically
convenient for the interacting electrons to 
localize at any lattice site. The energy gain associated
to the electronic motion becomes 
smaller than the cost associated to charge fluctuations. 
Hence, the commensurability of the electron density $n\!=\!N_{el}/N_{sites}$
naturally emerges as a necessary condition to obtain a Mott insulator.
Intuitively this commensuration implies the absence of spare sites
where an electron can hop without paying extra electrostatic energy. 

This idealized picture becomes less clear for more realistic systems, 
in which the correlated electrons in valence orbitals hybridize
with nearly uncorrelated ligand orbitals. 
This is the case of many different transition-metal  oxides (TMO) or
the Iron
pnictides/chalcogenides~\cite{Kamihara2008JACS,Yin2011NM,deMediciPRL2014,IronBook},
where the transition-metal $d$-orbitals are intertwined with
$p$-orbitals of the oxygen or pnictogen/chalcogen atoms. 
Investigations of these systems have suggested that a crucial
control parameter to determine the degree of correlation is the
filling of the correlated $d$
orbitals~\cite{DeMedici2009b,Weber2010PRB,Weber2010NP,Wang_Covalency,Parragh2013PRB,Hansmann2014NJOP,Razzoli_nd_correlations_pnictides}, 
which however does not generally correspond to an integer total number of electrons per site. 
Moreover, simplified models for these systems display seemingly contradictory results,
where the hybridization between a correlated electronic band and a wide band
of non-interacting electrons can either forbid~\cite{Medici2005PRL}  or allow~\cite{Georges1993,Sordi2007PRL,Amaricci2008PRL,Logan2016JOPCM}
the Mott transition in different physical regimes.

This rich spectrum of results leaves many questions open about the very presence of the Mott
transition in systems of this kind, and about the nature of the carriers which localize, i.e., if the 
ligand $p$ orbitals play a role, or they are mainly spectators of the $d$-electron localization.

In this work we provide an answer to some of these questions.
We study the conditions under which a Mott transition takes place, and what are its properties in a a simple, yet generic, model which 
captures the essential ingredients of the collective behavior of strongly correlated $d$-electrons 
hybridized with a band of non-correlated electrons. 
We solve the model using dynamical mean-field theory (DMFT)~\cite{Georges1996RMP,Kotliar2004PT}, a
powerful and reliable non-perturbative method which has been widely
used to address the Mott-Hubbard transition in a variety of models and
realistic systems~\cite{Kotliar2006RMP}. 
Our key result is to show that a zero temperature Mott  
transition exists only for an integer 
total density per site,
i.e. that the necessary condition is an integer total number of
electrons including also those occupying uncorrelated ligand orbitals.
We show that this transition corresponds to the localization of
singlets of mixed character between the two bands with density of one per lattice site. 
Finally, we characterize the nature of the correlated insulating
state with respect to experimentally relevant parameters of the
system.
We show that the mentioned past literature connects in this perspective and
fits well in the Zaanen-Sawatzky-Allen diagram~\cite{Zaanen1985PRL}.

\paragraph{Model. -- }
The starting point of our analysis is a generalized version of the
periodic Anderson model~\cite{Jarrell1993PRL,Rozenberg1995PRB,Jarrell1995PRB,
  Pruschke2000PRB,Sordi2007PRL,Amaricci2008PRL,Sordi2009PRB,Amaricci2012PRB}. 
The model Hamiltonian reads:
\eq{
\HH= \sum_{\ka\s} \psi^+_{\ka\s} \hat{h}(\ka)\psi_{\ka\s}
+ U\sum_i d^+_{i\up}d_{i\up} d^+_{i\dw}d_{i\dw}
}{Hpam}
where:
\eqnn{
\hat{h}(\ka) = 
\left(
\begin{array}{cc}
\ep0 + \e_\ka & \gamma(\ka) \\
\gamma^*(\ka)    & \ed0 + \a\e_\ka\\
\end{array}
\right)\, .
}

 For the sake of definiteness we consider a two-dimensional system on a square lattice.
The spinor $\psi_{\ka\s}\!=\!\pamrowvec^T$ collects the operators $p_{\ka\s}$, $d_{\ka\s}$ respectively annihilating electrons with spin $\s\!=\!\{\up,\dw\}$
in the wide band with dispersion
$\e_\ka\!=\!-2t_{pp}[\cos(k_x)+\cos(k_y)]$ centered at $\ep0$, and
in the narrow correlated band of dispersion $\a\e_\ka$
($\a\!\in\![0,1]$) centered at $\ed0$.
The two orbitals hybridize with an amplitude $\gamma(\ka)\! =\! \tpd -
4\v0\sin(k_x)\sin(k_y)$. 
The second term in (\ref{Hpam}) describes the local  
repulsion experienced by the $d$-electrons. 
In the following we set the energy unit to
the conduction electrons half-bandwidth $D\!=\!4t_{pp}\!=\!1$ and we introduce the charge-transfer energy
$\Delta\!=\!\ed0\!-\!\ep0$. We focus on the case
$\Delta\!\ge\! 0$ as similar results holds for the  $\Delta\!<\!0$ case. 
Without loss of generality we fix the zero of the energy to $\ed0\!=\!0$ and adjust the total density $\bra n \ket=\np\!+\!\nd\!$ (where $n_p\!=\sum_\s p^+_{ia\s}p_{ia\s}$ and $n_d\!=\sum_\s d^+_{ia\s}d_{ia\s}$ are the local number operators) by tuning the chemical potential $\mu$.
We consider the zero temperature regime $T=0$.

We solve the model (\ref{Hpam}) using Dynamical Mean-Field Theory (DMFT)~\cite{Georges1996RMP,Amaricci2012PRB}. We limit ourselves
to paramagnetic solutions in which magnetic ordering is neglected, in order to reveal the pure Mott physics. 
The lattice problem is mapped onto a quantum impurity problem
describing a single $d$-orbital coupled to 
an effective electronic bath~\footnote{An equivalent representation can be obtained considering a
  $p$-$d$ dimer embedded in a self-consistent bath, see \cite{Amaricci2012PRB}}. 
The auxiliary impurity problem is described in terms of the local {\it
  Weiss Field} (WF) $\GG_0^{-1}(\iome)$. 
A self-consistency condition relates the WF 
to the local properties of the lattice problem:
$\GG_0^{-1}\!=\!G_d^{-1}\! +\!\Sigma$,  
where $\Sigma$ is the self-energy function and $G_d$ is the
$d$-component of the local interacting Green's functions, i.e.
\eqsplit{
    G_d(\iome)  &= \sum_\ka \frac{\zeta_p-\e_\ka}{D_\ka(\iome)}\cr
    G_p(\iome)  &= \sum_\ka
    \frac{\zeta_d-\a\e_\ka}{D_\ka(\iome)}\cr
    D_\ka(\iome) & = 
    (\zeta_p-\e_\ka)(\zeta_d-\a\e_\ka)-|\gamma(\ka)^2|
  }{pamGFs}
with $\zeta_p=\iome\!+\!\mu\!-\!\ep0$ and $\zeta_d=\iome\!+\!\mu\!-\!\ed0\!-\!\Sigma(\iome)$.

The DMFT equations are closed by computing the impurity 
self-energy $\Sigma(\iome)$ from the numerical solution of the auxiliary impurity
problem~\cite{Georges1996RMP}.
In this work we use the exact diagonalization method, in which the
effective bath is discretized into a number $N_b$ of levels
\cite{Caffarel1994PRL}. The resulting Hamiltonian problem is solved
using the Lanczos algorithm to obtain both the lower part of the spectrum and
the Green's function \cite{Capone2007PRB,Weber2012PRB}. 
The results of this work have been obtained with $N_b=9$ and their
robustness with respect to $N_b$ has been tested in selected cases. 


\paragraph{Mott transition. -- }
In the symmetric case ($\Delta\!=\!0$, $\np\!=\!\nd\!=\!1$) the model 
describes a Kondo insulator for a large enough value of the 
local hybridization: $\tpd\!>\!\sqrt{\a}D$ \cite{Medici2005PRL} for
any interaction strength $U$. At smaller $\tpd$ a metal is obtained
instead with two bands of mixed $p$-$d$ character due to hybridization.
In the absence of hybridization the two bands
with different orbital character decouple and the $d$-band describes a half-filled Hubbard system. 
The latter is known to undergo a zero temperature Mott transition at a critical
interaction strength $U\!=\!U^\mathrm{HM}_{c2}$ of the order of the
three times the $d$-band half-bandwidth $\a D$. However, such Mott transition does not
survive the presence of any small hybridization~\cite{Medici2005PRL}.

\begin{figure}
\includegraphics[width=0.5\textwidth]{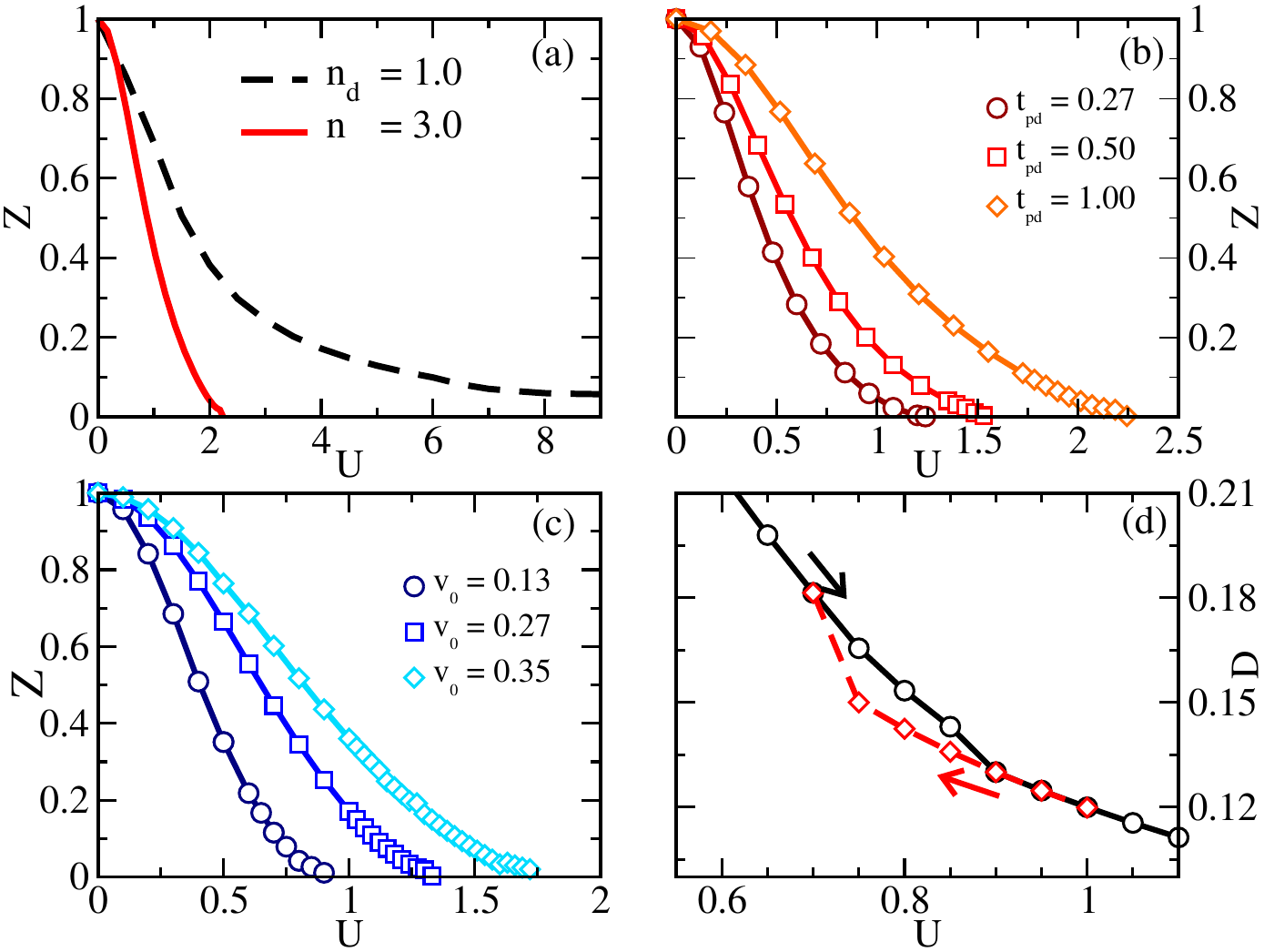}
\caption{(Color online) 
  Renormalization constant $Z$ and double occupancy $\bra d\ket$ of
  the correlated band for $\Delta\!=\!1.0$ and $\alpha\!=\!0.27$.
  (a) $Z$ as a function of the interaction $U$, for either
  $n_d\!=\!1.0$ (dashed line) or  $\bra n\ket\!=\!3.0$ (solid
  line). The other model parameter are  $\tpd\!=\!1.0$ and $\v0\!=\!0.0$.
  (b)-(c) $Z$ as a function of $U$ for $\tpd\!=\!0.0$ and increasing
  $\v0$ (b) or  $\v0\!=\!0.0$ and increasing $\tpd$ (c). 
  (d) Hysteresis cycle of the double occupancy  $\bra d\ket$
  near the Mott transition for $\v0\!=\!0.0$ and $\tpd=0.13$.
}
\label{fig1}
\end{figure}

Since no Mott transition can occur in the presence of hybridization in the 
symmetric model, we move to the mixed-valence regime ($\Delta\!>\!0$, $\nd\!\neq\!\np$), where the orbital occupations of the two orbitals are different.

To pinpoint a Mott transition, we compute the 
quasi-particle residue $Z=[1-\partial\Sigma/\partial\o]^{-1}$ as a
function of the interaction strength $U$ (black dashed line). The reduction to zero of this
quantity would signal a MIT. 
We first consider the case of integer filling for the
correlated orbitals only. In. \figu{fig1}(a) we present results where we actually fix $\nd\!=\!1$, 
which implies that the total density is different from an integer number and the $p$ bands 
are partially filled.
However, as our calculations show, 
$Z$ remains finite up to
huge values of the interaction, $U\!\gtrsim \!10D$.
In this regime the partially filled $p$ bands provide
a delocalization channel for the correlated electrons, 
ultimately preventing charge localization.

\begin{figure}
\includegraphics[width=0.5\textwidth]{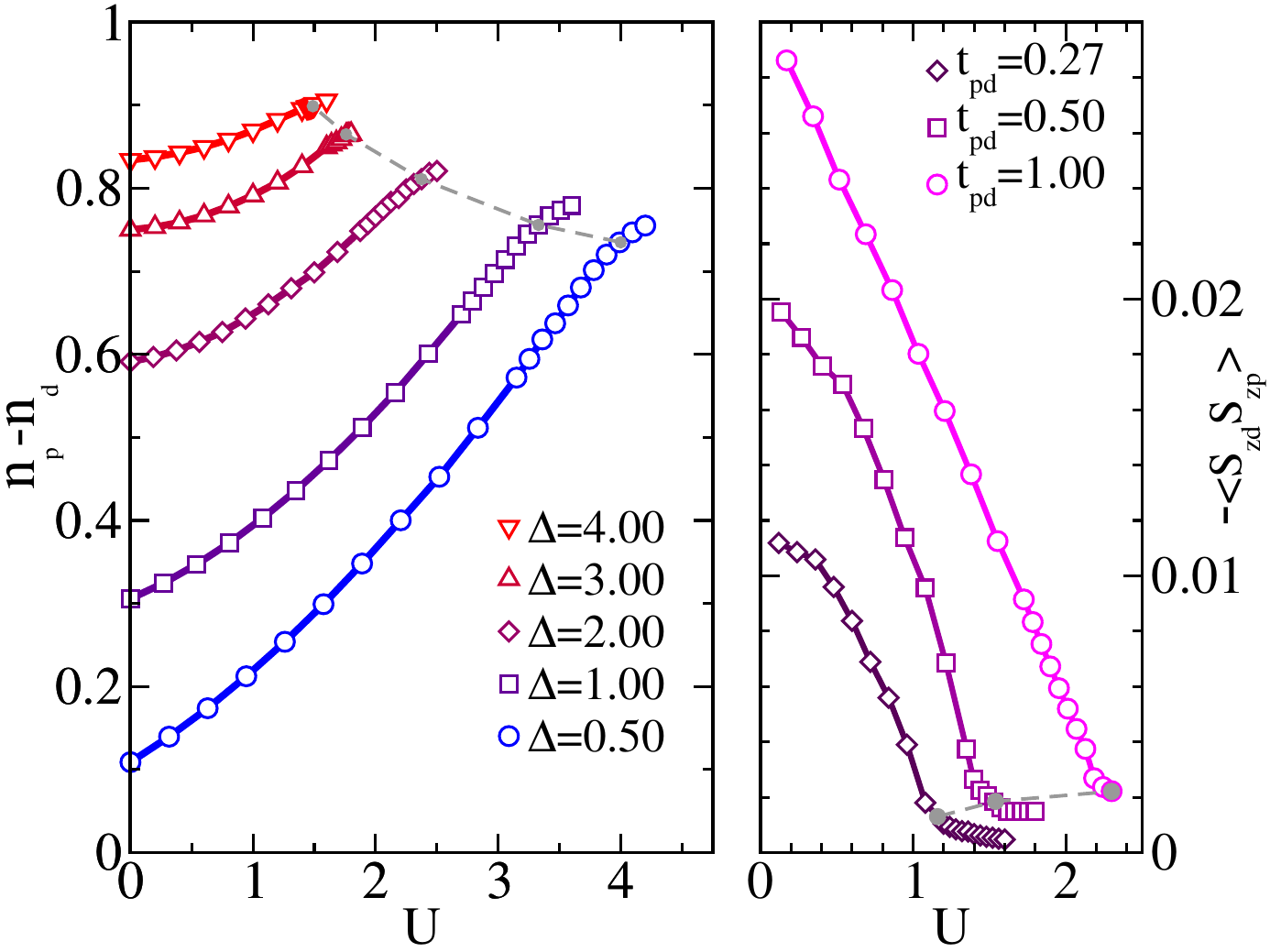}
\caption{(Color online) 
  {\it Left panel}: Orbital polarization
  $\np\!-\!\nd$ as a function of the $U$. Data are for increasing
  $\Delta$ and $\tpd=1.0$, $\v0=0.18$, $\a=0.27$. 
  {\it Right panel}:  Inter-orbital spin-spin correlation $\bra
  s_{dz}s_{pz}\ket$ as a function of $U$. Model parameters are as in
  \figu{fig1}(b). Grey dotted lines mark the Mott transition points.
}
\label{fig3}
\end{figure}

A Mott transition can however be realized in a different regime for
any value of the hybridization, as we shall show in the following.  
On rather general grounds, a sufficiently large hybridization drives the
formation of local singlets via the dynamical binding of two
electrons on different orbitals 
\cite{Zhang1988PRB,Sordi2009PRB} (see also \figu{fig3}). 
Thus, a MIT can be realized upon localization of
such composite fermionic states for large enough interaction, provided their
occupation is one per lattice site. We now show that this is  realized  for a 
total density  $\bra n\ket\!=\!\nd+\np=\!3$.
In \figu{fig1}(a) we report the behavior of $Z$ for this
occupation. Our results show that $Z$ approaches zero at a critical value of the
correlation $U=U_{c2}$. 
The existence of a Mott transition in this regime turns out to be generic with
respect to the amplitude and the character of the hybridization, as shown in \figu{fig1}(b)-(c), where
we compare results for  local and non-local hybridizations. In both cases we observe the existence of 
a metal to Mott insulator transition at a critical interaction increasing
with the hybridization amplitude. 
Similarly to the single-band Hubbard model, we find a coexistence of metallic and insulating solutions. This is
demonstrated by our results in \figu{fig1}(d), reporting a small
hysteresis cycle of the double occupation $\bra d\ket\! =\! \bra
n_{d\up} n_{d\dw}\ket$ at the $d$-orbital. 

To better understand the mechanism behind the MIT at $\bra n\ket\!=\!\nd+\np=\!3$ and the absence of a transition 
for $\nd\!=\!1$, we follow the evolution of the orbital occupations 
$\bra n_p\ket$ and $\bra n_d\ket$.  
Our results are summarized in the left panel of Fig.~\ref{fig3}, which reports the
orbital polarization $\np\!-\!\nd$ as a function of the correlation $U$. 
In the integer-valence regime, where a pure-$d$ picture would applies,
$\bra n_p\ket=2$ and $\bra n_d\ket=1$. 
As shown by our calculations, the system remains in its
intermediate-valence regime up and throughout to the transition, i.e. 
the Mott transition occurs with a
non-integer value of the individual occupations of the orbitals.
Increasing the interaction strength $U$ enhances the orbital
polarization, by transferring charge from the correlated orbital to the
uncorrelated orbitals. 
%
For a fixed value of $U$, the orbital polarization
naturally increases as a function of the charge-transfer energy $\Delta$. This
effect translates the fact that orbital mixing is larger when the
two orbitals are brought near one another in energy, and decreases rapidly with
increasing energy separation. Indeed, the integer-valence regime
can only be reached  asymptotically in the
$\Delta\!\rightarrow\!\infty$ limit~\cite{Sordi2009PRB}. 

In the right panel of \figu{fig3}, we report the evolution of
inter-orbital spin-spin correlation $\bra s_{zd}s_{zp}\ket$. 
This quantity describes the binding of the 
magnetic moments between the $d$- and $p$-electrons, which defines the
formation of  the local singlets.  
Our results show that the inter-orbital spin-spin correlation remains
finite across the MIT, exhibiting  the persistence of the
local singlets in the Mott state. 
The  decreasing behavior of  $\bra s_{zd}s_{zp}\ket$ as a function of
$U$ results from the renormalization of the hybridization
amplitude, which ultimately leads to a loosening of the magnetic
binding.

\begin{figure}
\includegraphics[width=0.5\textwidth]{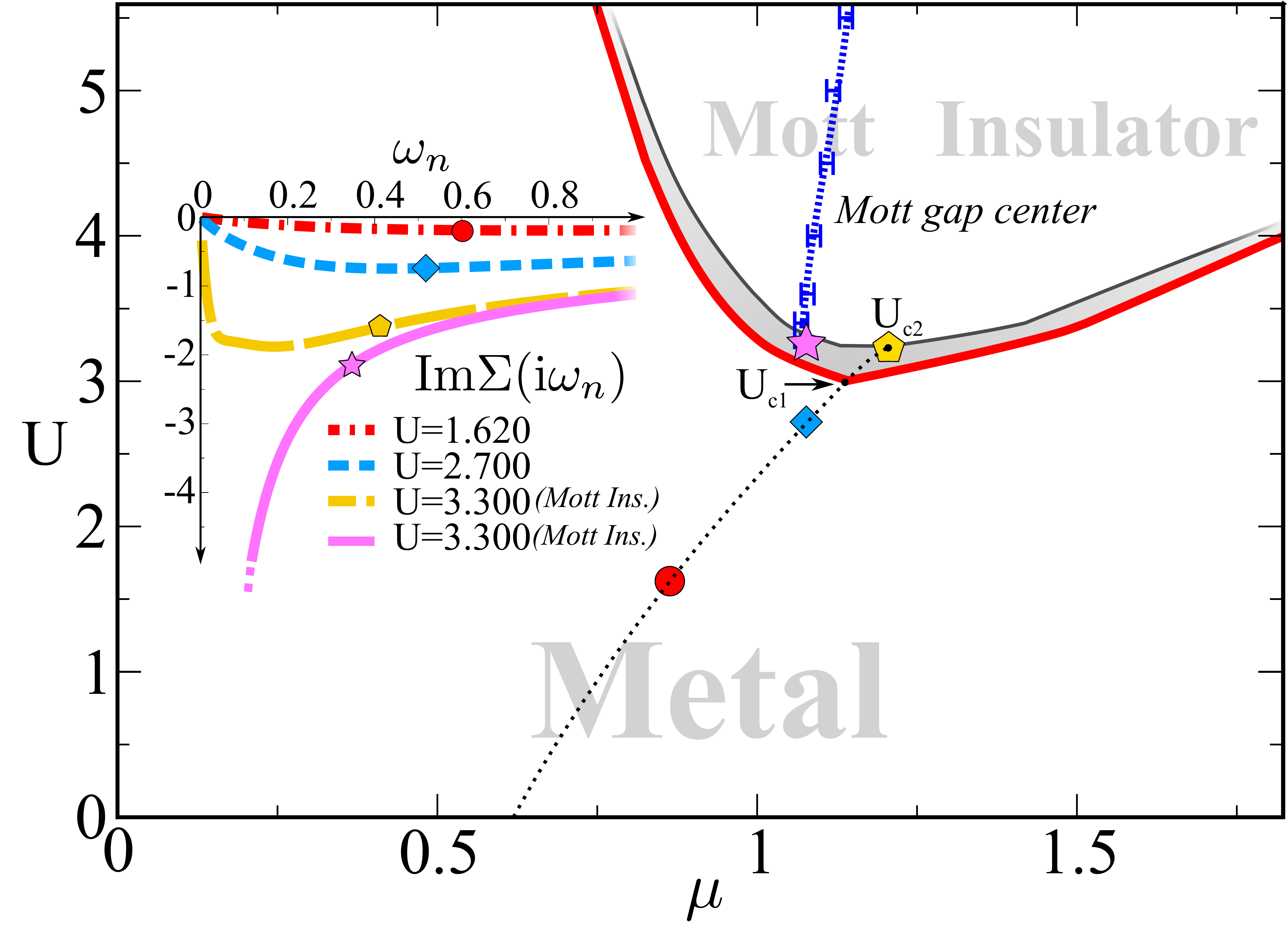}
\caption{(Color online) Phase-diagram in the  $U$-$\mu$ plne. Data
  are for $\tpd\!=\!1.0$, $\v0\!=\!0.18$, $\a\!=\!0.27$ and
  $\Delta\!=\!1.0$.  
  The (red) solid line is the boundary of the Mott region. 
  The shaded (gray) area indicates the coexistence
  regions of metallic and insulting solution. 
  The (black) dotted line indicates the $\bra n\ket\!=\!3$ path,
  ending at $U\!=\!U_{c2}$. 
  The symbols mark the points corresponding to  $\Im\Sigma(i\omega_n)$
  curves in the inset. The (blue) dashed line and points indicate
  center of symmetry for the Mott gap (see main text).  
}
\label{fig2}
\end{figure}

\paragraph{Mott gap opening. -- } 
In order to investigate the opening of the Mott gap in 
the intermediate-valence regime, we mapped out the $U$-$\mu$ phase
diagram, reported in \figu{fig2}. The condition $\bra n\ket\!=\!3$
determines a line in the metallic part of the diagram intercepting  the ``V''-shaped
Mott insulating region at $U\!=\!U_{c1}$. The metallic solution however disappears
inside the insulating region at the critical $U\!=\!U_{c2}\!>\!U_{c1}$ point. 
The opening of the Mott gap is associated to the presence of a pole in the self-energy on the real-axis $\Im\Sigma(\omega\!+\!i\eta)$ close to ($\omega\!=\!0$). 
When the pole is not exactly at zero frequency, the large value of 
the real part of the self-energy prevents quasiparticle
excitations inside the gap.  Correspondingly (due to the Kramers-Kr\"onig relations) the imaginary part 
 is found to go to zero at $\omega\!=\!0$, i.e. $\Im\Sigma(i\omega_n)$ vanishes on the {\it Matsubara} axis with
a very large linear slope (see inset of \figu{fig2}). 
The set of points in the diagram where $\Im\Sigma(i\omega_n)$
instead diverges, i.e. when the pole of the self-energy is at $\omega\!=\!0$,
determine  the center of symmetry of the Mott gap (see dotted line in
\figu{fig2}).

\begin{figure}
\includegraphics[width=0.49\textwidth]{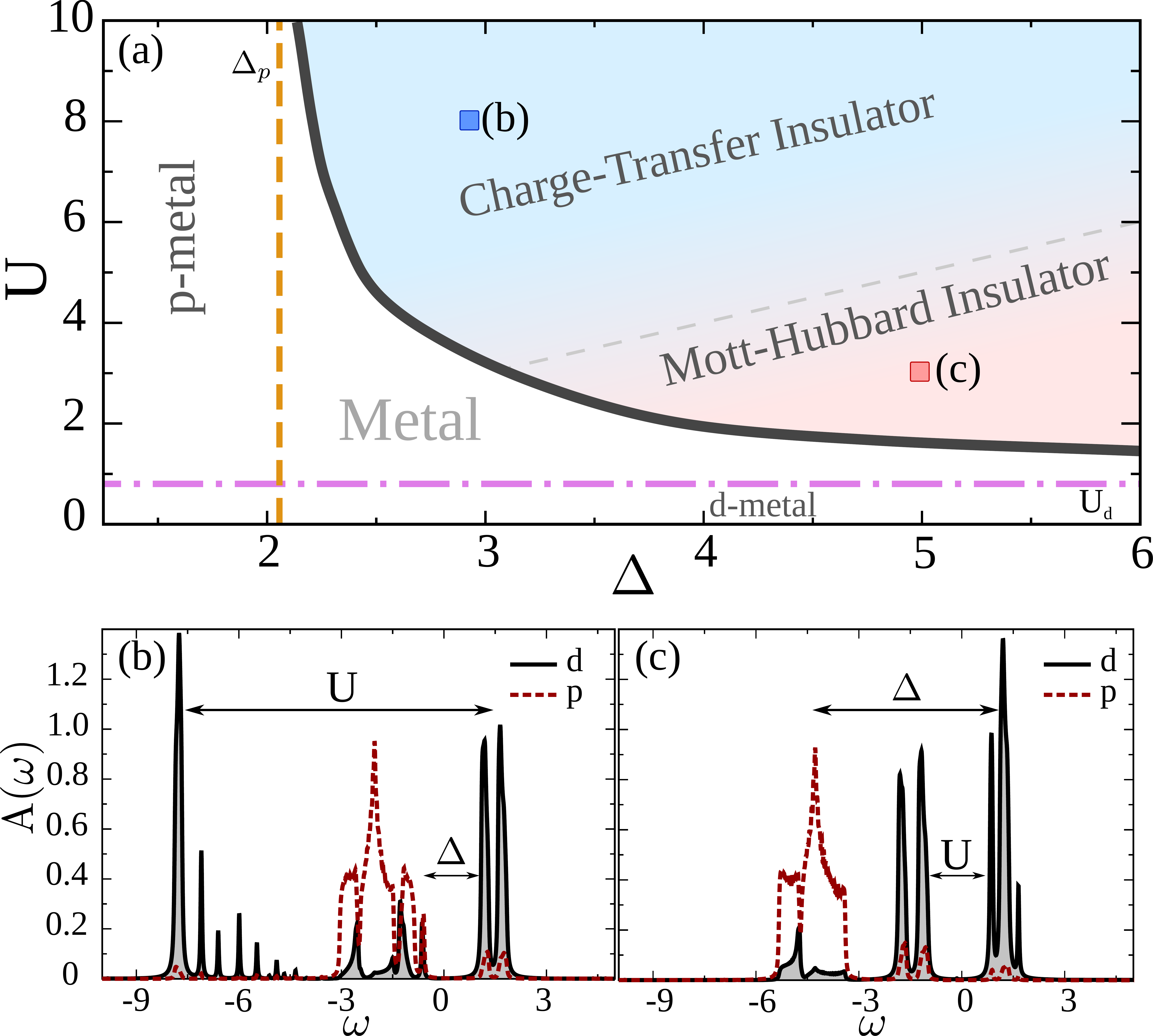}
\caption{(Color online) 
  (a) Phase-diagram in the $U$-$\Delta$ plane, for $\alpha\!=\!0.27$,
  $\tpd\!=\!1.0$ and $\v0\!=\!0.18$. The (black) solid line delimitates
  the insulating region, separated from the  metallic solution by a
  first order transition. 
  The insulating region is divided in two parts: the Charge-Transfer
  (b) and the Mott-Hubbard (c) for (see main text). 
  The lower panels show the spectral densities 
  $A_\alpha(\omega)\!=\!-\Im G_{\a}(\omega)/\pi$, $\a\!=\!d,p$, for
  the two points (b),(c) in the diagram.
  The vertical line at $\Delta\!=\Delta_p\!=\!2.1$ is the boundary of
  the $p$-metal regime, while the horizontal line at
  $U\!=\!U_d\!=\!1.3$.delimitates the $d$-metal (see main text). 
}
\label{fig4}
\end{figure}

\paragraph{Phase-diagram. -- }
Finally, we studied the dependence of the Mott transition with respect
to the charge-transfer energy $\Delta$. 
Our findings are reported in \figu{fig4}(a), showing the
phase-diagram in the  $U$-$\Delta$ plane, which closely follows the Zaanen-Sawatzky-Allen diagram~\onlinecite{Zaanen1985PRL}.
We identify two distinct regimes: A charge-transfer insulator (CTI) for $U\!\gtrsim\!\Delta$ and
a Mott-Hubbard insulator (MHI) for $U\!\lesssim\!\Delta$. 
The different nature of these two insulating solutions is underlined
by the spectral functions $A_{p/d}(\omega)=-\Im
G_{p/d}(\omega)/\pi$, presented in the bottom panels of the figure. 
For the CTI (see \figu{fig4}(b)) the smallest gap is of order $\Delta$
and corresponds to the excitations from the wide central band, with
prevalently $p$-character, to the upper Hubbard band. 
Conversely, in the MHI (see \figu{fig4}(c)) the smallest gap is set by the
energy separation of the two Hubbard band of order $U$. 
A small coexistence region of metallic and insulating solutions is found.
The boundary line for the Mott transition is proportional to $1/\Delta$. 
For large $\Delta$ the boundary line approaches the 
critical value of a Hubbard system for  the
correlated band only, i.e. $U_d\!=\!\a U_{c2}^\mathrm{HM}$ ($d$-metal). 
This corresponds to the suppression of the indirect 
delocalization  of the correlated electrons of the order of
$\sum_\ka |\gamma(\ka)|^2/\Delta$. 
In the limit of large interaction $U$ a MIT is obtained by closing the charge-transfer gap. 
For any finite hybridization, there exist a critical
charge-transfer energy  $\Delta_p$ below which the
conduction band crosses the Fermi level. As follows from the structure
of the Eqs.~\ref{pamGFs},  the critical value $\Delta_p$ is ultimately
determined by the equation:
$D_\ka(0)\!=\!0$, 
i.e. $\e_\ka\!+\!\ep0\!-\!\mu\!=\!\tfrac{\gamma(\ka)}{\alpha\e_\ka\!+\!\Re\Sigma(0)\!-\!\mu}\,.$ 
which corresponds to having a Green's functions pole at the Fermi level,
for any value of the interaction ($p$-metal). The solution of this
equation is possible for any $Re\Sigma(0)$ as long as the bare p-band
crosses the Fermi level. The absence of this band at the Fermi level
is thus a necessary condition for the Mott transition. Since for any
fixed hybridization the charge-transfer energy tunes monotonically the
density of correlated electrons $\nd$, this motivates the arising of a
non-integer threshold level of $\nd$ for the Mott transition at large
U, as reported in \cite{Wang_Covalency} and illustrated by the dashed
line in \figu{fig3}a. This value depends on the
other parameters (most notably the hybridization) and is
non-universal.
Our conclusion is thus that the integer criterion for the total
density plus the requirement that the bare $p$-band does not crosses the
Fermi level rationalizes the seemingly diverging results of the
literature cited in the introduction, under the general perspective of
the Zaanen-Sawatzky-Allen diagram~\cite{Zaanen1985PRL}.

\paragraph{Conclusions. -- }
We investigated a paradigmatic model of strongly correlated
electrons hybridized with  a non-interacting ligand band. 
We demonstrated that a $T=0$ Mott transition 
occurs for any non-vanishing hybridization amplitude in the 
mixed-valence regime. This transition corresponds to the Mott localization
of singlet states formed by the binding between correlated and
conduction band electrons. 
We point out that a necessary condition for the Mott transition 
is to have an odd total integer filling $\bra n\ket\!=\!3$,
corresponding to a singlet density of one per site but to a
non-integer filling of the correlated orbitals. 
Finally, we identify the first-order nature of the transition and
discuss the mechanism for the gap opening as well as its dependence on
the charge-transfer energy, pointing out that MIT only occurs  if the bare
$p$-band does not cross the Fermi level.
These results are relevant to understand the physics of $d$-orbital materials,
such as TMO.  
The extension of this study to the more realistic case of
multi-orbital systems, which better capture the physics of $t_{2g}$
and $e_g$ orbitals and their different hybridization with ligand atoms in TMO, is an interesting future direction of research. 

\paragraph*{Acknowledgments.} 
The authors are indebted with M.~J.~Rozenberg, G.~Sangiovanni and A.~Valli for useful discussion.
Financial support from the European Research
Council under FP7 Starting Independent Research Grant n.240524
``SUPERBAD", Seventh Framework Programme FP7, under Grant No. 280555 ``GO FAST''
and under H2020 Framework Programme, ERC Advanced Grant No. 692670
``FIRSTORM'' are acknowledged. 
\bibliography{localbib}
\end{document}